\documentclass[journal]{IEEEtran}
\usepackage{cite}
\usepackage{amsmath,amssymb,amsfonts}
\usepackage{graphicx} 
\usepackage{textcomp,nicefrac}
\usepackage{booktabs}

\usepackage[switch]{lineno}
\begin{document}
\title{Scintillation Properties of a Stilbene Crystal for Low-Mass Dark Matter Searches}

\author{Se~Hwan~Lee, Jaeyoung~Cho, D. Joseph Daniel, Hongjoo~Kim, Young~Ju~Ko, Jungho~So, and~In~Soo~Lee
\thanks{S. H. Lee and Y. J. Ko are with Jeju National University, Jeju, Korea.}
\thanks{J. Cho, D. Joseph Daniel, and H. Kim are with Kyungpook National University, Daegu, Korea.}
\thanks{I. S. Lee is with the Center for Underground Physics, Institute for Basic Science (IBS), Daejeon, Korea (corresponding author, e-mail: islee@ibs.re.kr).}
\thanks{J. So is with the Yemilab Operation Center, Institute for Basic Science (IBS), Jeongseon, Korea.}
}

\maketitle

\begin{abstract}
Direct searches for low-mass WIMPs require sensitivity to low-energy nuclear recoils. Hydrogen-containing targets offer favorable scattering kinematics because low-mass WIMPs can transfer a larger fraction of their kinetic energy to hydrogen nuclei than to heavier nuclei. Trans-stilbene (t-stilbene) is a hydrogen-rich organic crystal that combines this kinematic advantage with efficient scintillation and pulse-shape discrimination (PSD).

In this work, we characterize the scintillation decay behavior and effective light yield of a solution-grown t-stilbene crystal to evaluate its suitability for dark matter detection. The detector consists of a cylindrical crystal approximately 1.1~cm in diameter and 1.1~cm in length, optically coupled at opposite ends to two Hamamatsu R12669 photomultiplier tubes. The scintillation waveforms are described by a triple-exponential decay model, yielding fraction-weighted decay time constants of $8.3 \pm 0.2$~ns (fast), $20.1 \pm 0.3$~ns (medium), and $74.2 \pm 1.1$~ns (slow). The effective light yield, determined using a 59.54~keV $\gamma$ ray from an $^{241}$Am source, is $3.37 \pm 0.02$ photoelectrons per keV.

These measurements establish the baseline performance of the detector and support further evaluation of t-stilbene as a target material for rare-event and low-mass dark matter searches.
\end{abstract}

\section{Introduction}
\label{sec:introduction}
\IEEEPARstart{T}{he} direct detection of weakly interacting massive particles (WIMPs), a leading class of dark matter candidates, remains a major challenge in particle physics and astrophysics \cite{pdg2022}. Although large-scale liquid-noble-gas detectors have set stringent limits on the WIMP--nucleon scattering cross section for masses above a few tens of GeV/$c^2$ \cite{xenonnt2023, lz2023, darkside2023}, growing theoretical and experimental interest is focused on the low-mass regime, from sub-GeV masses to a few GeV/$c^2$. Searches in this regime require detector materials with very low energy thresholds and effective background discrimination at low recoil energies.

Trans-stilbene offers several advantages for low-mass dark matter searches. First, its hydrogen-rich composition can produce higher observable recoil energies than those obtained with heavier target nuclei, such as xenon or iodine, thereby improving sensitivity at low energy thresholds~\cite{battaglieri2017us}. Second, its fast timing response and pulse-shape discrimination (PSD) capability, which is associated with delayed fluorescence from triplet--triplet annihilation, can help distinguish electron-recoil backgrounds from WIMP-induced nuclear recoils \cite{knoll2010}. Stilbene also exhibits directional anisotropy: the proton-recoil quenching factor depends on the crystal axis \cite{shimizu2003}. This property could be used to search for the directional signature of the WIMP wind and provide an additional handle for background rejection \cite{shimizu2003}.

A dedicated dark matter search based on t-stilbene requires a high-purity crystal with well-characterized detector performance. The t-stilbene single crystal investigated in this study was produced using a solution-growth method. We performed a baseline characterization of the crystal for rare-event applications, focusing on its effective light yield and scintillation decay behavior.

This article presents a characterization of a t-stilbene crystal detector assembly. We evaluate its optical and scintillation properties, including the photoluminescence spectrum, scintillation decay constants, and effective light yield determined through a single-photoelectron calibration.

\section{Experimental Setup}
\label{sec:experimental}

\subsection{Crystal Growth}
The t-stilbene single crystal used in this study was grown by a controlled solution-growth method. A circular seed crystal with a thickness of 5 mm was prepared by conventional slow evaporation. The seed was oriented with its broad surface parallel to the (001) crystallographic plane and was placed horizontally at the bottom of the ampoule. A saturated, homogeneous stilbene solution was maintained at $45^\circ\mathrm{C}$, and a few drops of pure xylene were added to suppress secondary nucleation during filtration and transfer. The solution was filtered through Whatman filter paper using a motorized pumping system and then carefully transferred into the growth ampoule without disturbing the seed.

The suspended ampoule was vertically oriented and axially symmetric, providing stable conditions for unidirectional crystal growth. The solution temperature was first increased from $45^\circ\mathrm{C}$ to $50^\circ\mathrm{C}$, slowly reduced to $45^\circ\mathrm{C}$, and then held at this temperature for 5~h. The bath temperature was regulated by a programmable controller with an accuracy of $\pm0.01^\circ\mathrm{C}$. A high-quality single crystal was obtained using an optimized cooling rate of $0.4^\circ\mathrm{C}/\mathrm{day}$. As the solution became supersaturated during continuous cooling, solute molecules were gradually incorporated into the exposed surface of the seed. This stable, undisturbed process promoted uniform single-crystal growth.

\begin{figure*}[!htbp]
  \centering
  \begin{tabular}{@{}c@{\hskip 0.005\textwidth}c@{\hskip 0.005\textwidth}c@{}}
    \includegraphics[width=0.31\textwidth,height=4cm,keepaspectratio]{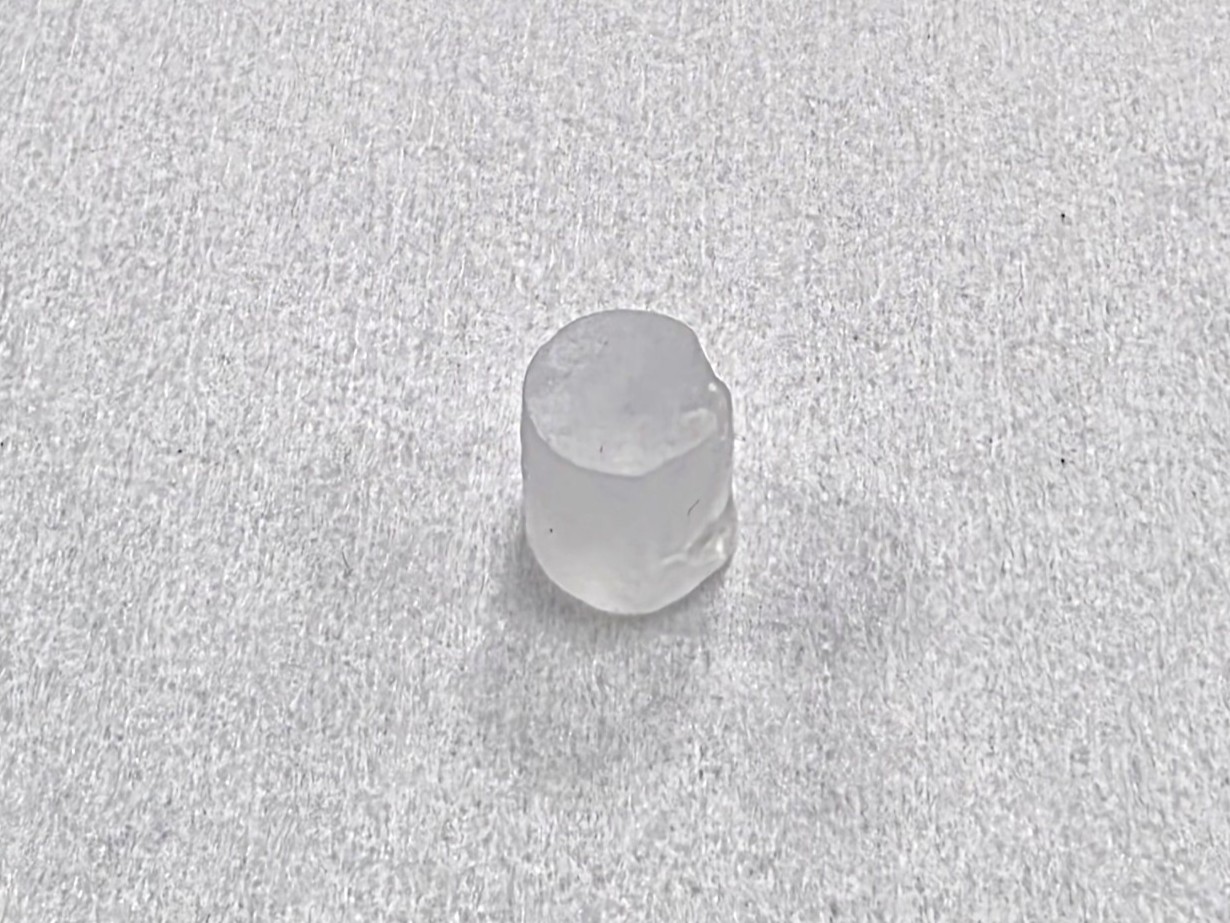} & 
    \includegraphics[width=0.31\textwidth,height=4cm,keepaspectratio]{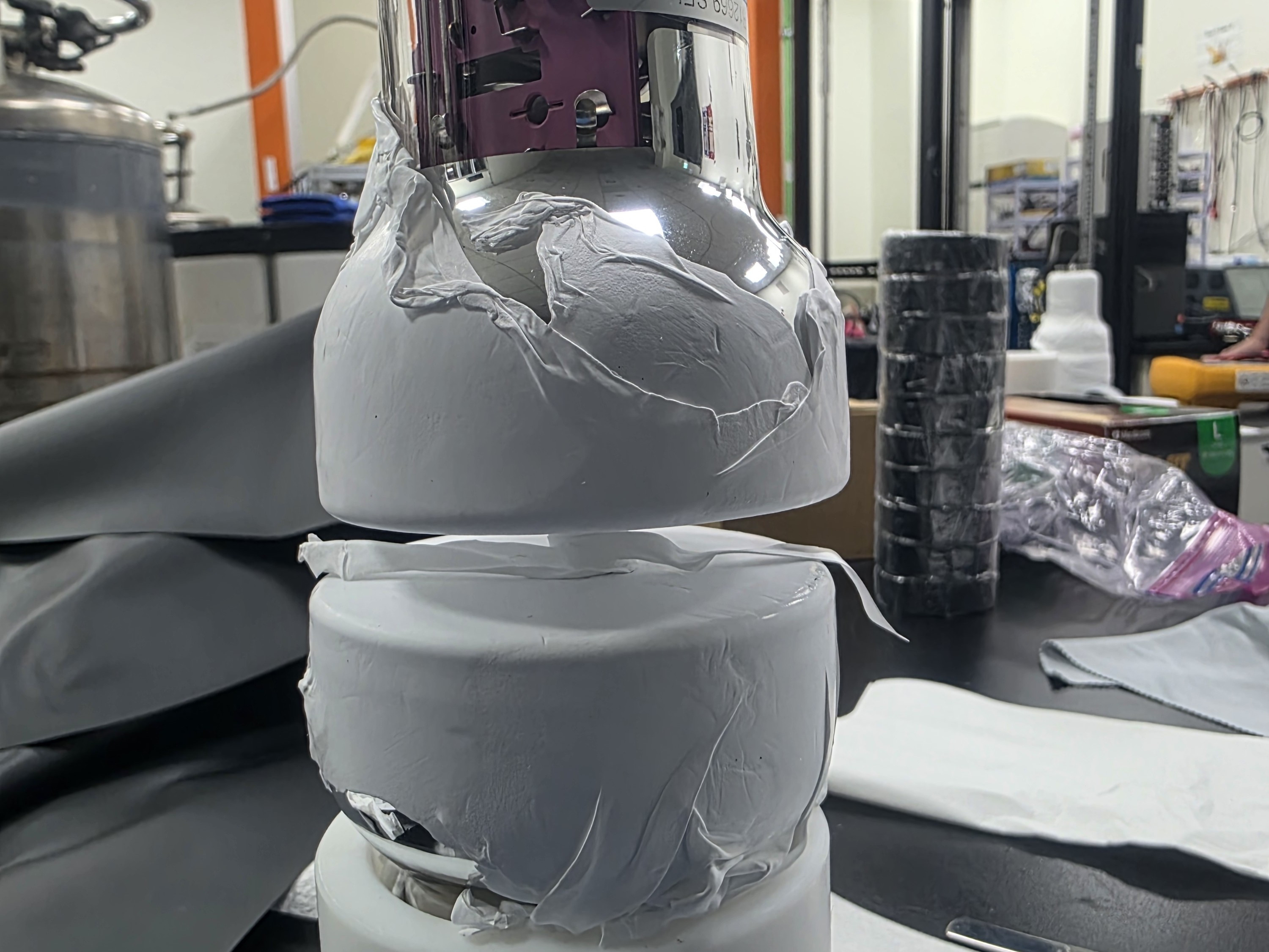} & 
    \includegraphics[width=0.31\textwidth,height=4cm,keepaspectratio]{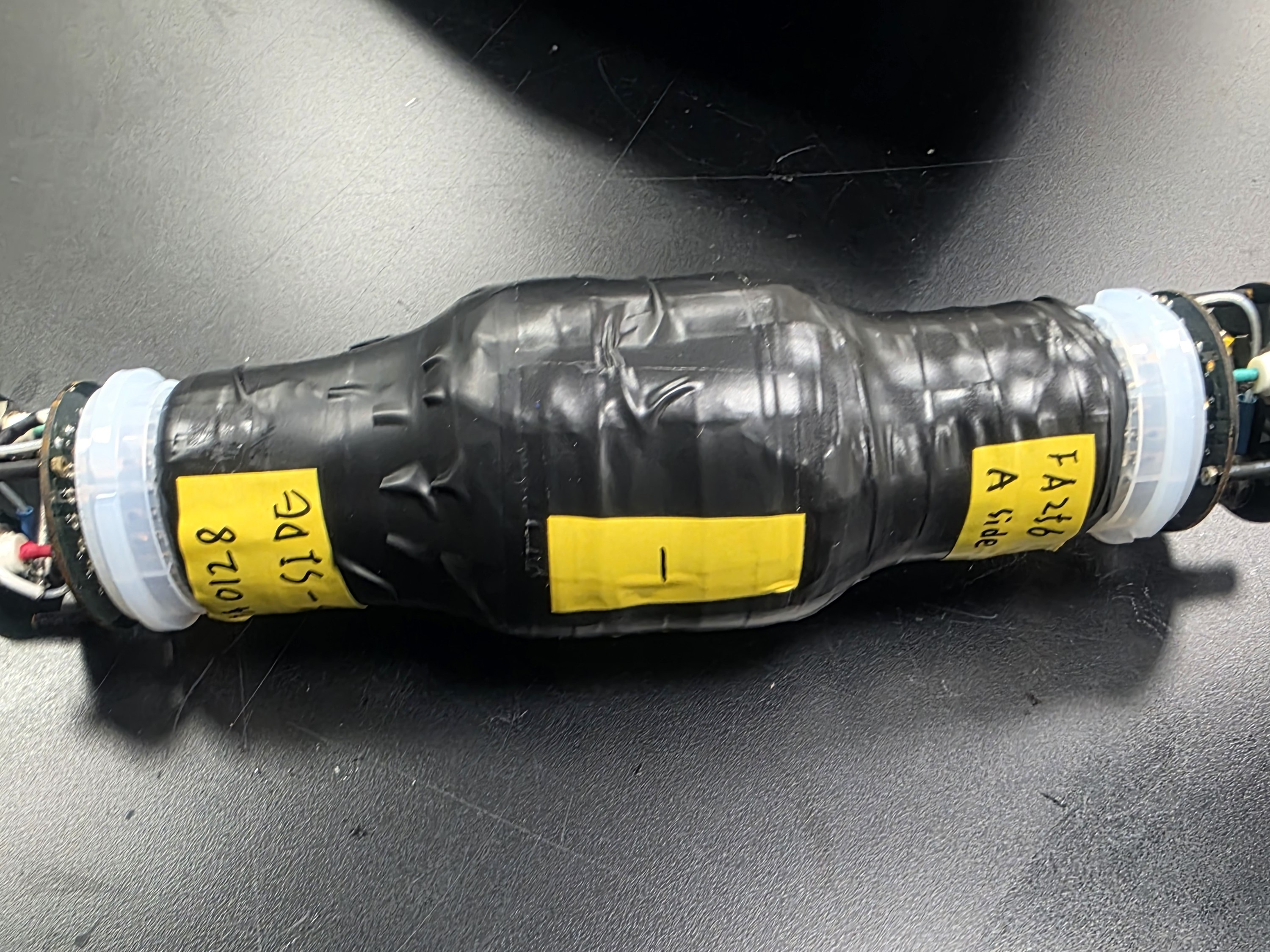} \\
    (a) & (b) & (c)
  \end{tabular}
  \caption{Preparation of the stilbene detector assembly: (a) bare cylindrical t-stilbene crystal, approximately 1.1~cm in diameter and 1.1~cm in length; (b) crystal wrapped in PTFE tape and coupled to the PMTs; and (c) completed detector assembly sealed with black tape to exclude ambient light.}
  \label{fig:crystal}
\end{figure*}

\subsection{Detector Configuration and Shielding}
The detector used a cylindrical t-stilbene crystal cut from the solution-grown crystal described above. The sample was approximately 1.1~cm in diameter and 1.1~cm in length, as shown in Fig.~\ref{fig:crystal}. The two flat faces were optically coupled with grease to 3-inch Hamamatsu R12669 photomultiplier tubes (PMTs). The crystal was wrapped in a PTFE (Teflon) reflector to improve light collection, and black tape was applied over the assembly to maintain mechanical contact and exclude ambient light.

\begin{figure}[htbp]
  \centering
  \includegraphics[width=0.8\linewidth]{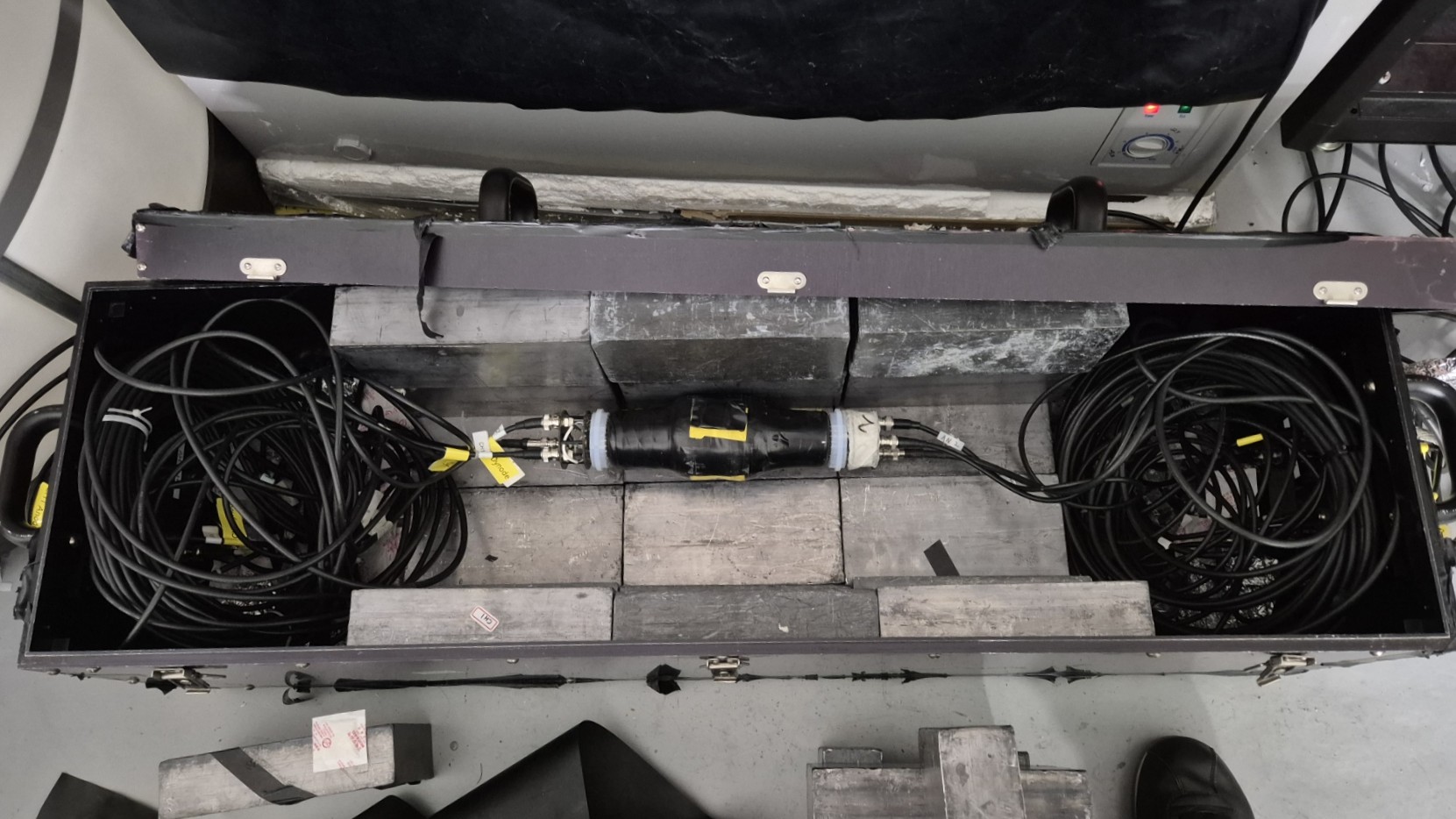}
  \caption{Experimental shielding setup, with the stilbene detector assembly placed inside a dark box surrounded by lead bricks.}
  \label{fig:shielding}
\end{figure}

To reduce external radiation backgrounds, the detector was placed in a dark box surrounded by 10-cm-thick lead shielding, as shown in Fig.~\ref{fig:shielding}. During data acquisition, additional lead bricks were placed above the closed dark box, and the full setup was covered with an opaque black sheet.

\subsection{Data Acquisition System}
The analog output from each PMT was amplified by a factor of 30 using a preamplifier and then digitized with a 12-bit, 500~MS/s FADC-500 digitizer (Notice Korea) \cite{lee2018}.

Events were recorded when signals from PMT1 and PMT2 occurred within a 200~ns coincidence window. This trigger condition suppressed uncorrelated PMT noise while retaining scintillation events in the crystal. Each waveform spanned 8.16~$\mu$s, with the trigger located 2.4~$\mu$s from the beginning of the acquisition window, providing sufficient post-trigger coverage for the full scintillation pulse and its late-time tail.

\section{Results and Discussion}
\label{sec:results}

\subsection{Optical Properties}
\begin{figure}[htbp]
  \centering
  \includegraphics[width=\linewidth]{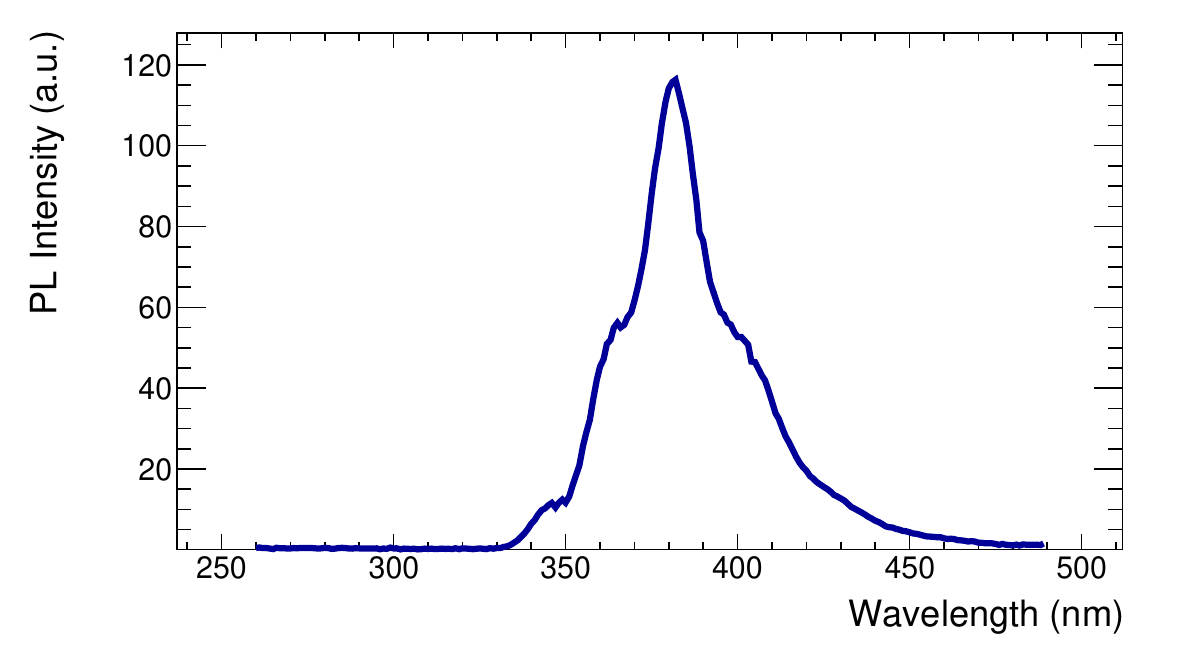}
  \caption{Photoluminescence emission spectrum of the t-stilbene crystal under 250~nm excitation, measured with a Cary Eclipse fluorescence spectrophotometer~\cite{cary_eclipse}. The spectrum shows a prominent emission peak near 380~nm.}
  \label{fig:pl}
\end{figure}

The optical emission properties of the t-stilbene crystal were evaluated using photoluminescence (PL) measurements. The PL spectrum was measured at room temperature with a Cary Eclipse fluorescence spectrophotometer (Agilent Technologies) \cite{cary_eclipse}. The sample was excited at 250~nm, and the emission spectrum was recorded from 300 to 500~nm. As shown in Fig.~\ref{fig:pl}, the spectrum exhibits a prominent peak near 380~nm, consistent with previously reported values \cite{zaitseva2015, inrad_stilbene}.

\subsection{Scintillation Decay Time Analysis}
\label{subsec:decay}

\begin{figure*}[!htbp]
  \centering
  \includegraphics[width=\linewidth]{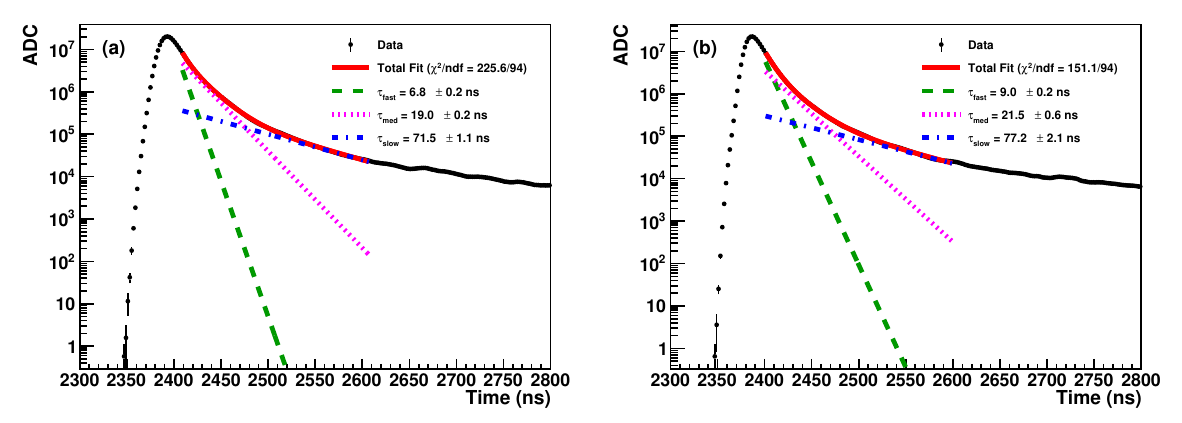}
  \caption{Acquired scintillation waveforms of the t-stilbene crystal, corresponding to the 59.54~keV $\gamma$-ray full-energy peak from an $^{241}$Am source, fitted with a triple-exponential model for (a) PMT1 and (b) PMT2. Data points are shown as black circles. The total fit is represented by the solid red line, comprising the fast (green long-dashed), medium (magenta dotted), and slow (blue dash-dotted) decay components.}
  \label{fig:decay}
\end{figure*}

Characterizing the time profile of the scintillation signal is important for constructing pulse-shape models and optimizing charge-integration windows. Data were collected for approximately 13~h with the t-stilbene crystal exposed to an $^{241}$Am source. Events within an approximately $3\sigma$ region around the 59.54~keV $\gamma$-ray full-energy peak were selected. Their waveforms were summed without aligning the individual pulse maxima, thereby preserving the measured timing distribution of the detector system.

As shown in Fig.~\ref{fig:decay}, the summed waveform from each PMT was fitted with a triple-exponential model. The fit range began slightly after the nominal trigger position (approximately 2400~ns) to exclude possible trigger-related or electronic artifacts and isolate the decay region. The fit function was
\begin{equation}
\label{eq:decay}
L(t) = \sum_{i=1}^{3} A_i \exp\left(-\frac{t - t_0}{\tau_i}\right),
\end{equation}
where $\tau_i$ denotes the decay time constant of the fast, medium, or slow component, $A_i$ is its initial amplitude, and $t_0$ is the effective start time of the fitted decay.

The fractional contribution of each component was obtained by integrating the corresponding exponential term; the integrated area is proportional to $A_i\tau_i$. Because the two PMTs showed modest differences in their temporal responses, possibly owing to light-collection asymmetry, PMT transit-time differences, or unequal cable lengths, the decay parameters were fitted separately. For PMT1 (PMT2), the extracted time constants were $6.80 \pm 0.22$~ns ($8.99 \pm 0.22$~ns) for the fast component, $19.04 \pm 0.22$~ns ($21.54 \pm 0.56$~ns) for the medium component, and $71.51 \pm 1.09$~ns ($77.24 \pm 2.10$~ns) for the slow component. The corresponding fractional contributions for PMT1 (PMT2) were $17.5 \pm 0.8\%$ ($36.6 \pm 1.8\%$), $64.8 \pm 2.0\%$ ($48.3 \pm 3.2\%$), and $17.7 \pm 0.8\%$ ($15.1 \pm 1.2\%$), respectively. Fraction-weighted values for the detector assembly are summarized in Table~\ref{tab:decay}.

\begin{table}[htbp]
\centering
\caption{Scintillation Decay Time Constants and Fractional Contributions}
\label{tab:decay}
\resizebox{\columnwidth}{!}{%
\begin{tabular}{@{}lcccc@{}}
\toprule
\textbf{Component} & \textbf{PMT1 $\tau$ [ns]} & \textbf{PMT2 $\tau$ [ns]} & \textbf{Fraction-Weighted $\tau$ [ns]} & \textbf{Mean Fraction [\%]} \\ \midrule
Fast               & $6.8 \pm 0.2$             & $9.0 \pm 0.2$             & $8.3 \pm 0.2$                & $27.3 \pm 0.9$              \\
Medium             & $19.0 \pm 0.2$            & $21.5 \pm 0.6$            & $20.1 \pm 0.3$               & $56.4 \pm 2.0$              \\
Slow               & $71.5 \pm 1.1$            & $77.2 \pm 2.1$            & $74.2 \pm 1.1$               & $16.4 \pm 0.7$              \\ \bottomrule
\end{tabular}%
}
\end{table}

As summarized in Table~\ref{tab:decay}, the three-component model provides a consistent qualitative description of the signals from both PMTs, although the fitted parameters differ. The medium component, with a fraction-weighted decay time of 20.1~ns, accounts for most of the emitted light. The fast component (8.3~ns) describes the prompt decay after the pulse maximum, whereas the slow component (74.2~ns, 16.4\% of the integrated signal) is consistent with delayed fluorescence associated with triplet-exciton annihilation.

The measured fast decay time constant of 8.3~ns is longer than the value of approximately 4.3~ns reported for pure t-stilbene crystals \cite{inrad_stilbene, harihar1993}. This difference can be attributed, at least in part, to the instrumental response of the data acquisition system. The 500~MS/s sampling rate corresponds to a 2~ns sample spacing, which limits the characterization of very fast signals. The preamplifier and signal cables may introduce additional pulse broadening. The measured 8.3~ns value should therefore be interpreted as an effective decay constant that includes the convolution of the intrinsic scintillation response with the electronics response.

\subsection{Effective Light Yield Measurement}
\label{subsec:ly}
The effective light yield is a key parameter in determining the achievable energy threshold of a dark matter detector. We measured the light yield in photoelectrons per keV (NPE/keV) using the 59.54~keV $\gamma$ ray emitted by an $^{241}$Am source.

The light yield ($LY$) was calculated as
\begin{equation}
\label{eq:ly}
LY (\text{NPE/keV}) = \frac{Q_{^{241}\text{Am}}}{Q_{SPE}} \times \frac{1}{E_{^{241}\text{Am}}},
\end{equation}
where $Q_{^{241}\text{Am}}$ is the total charge, in ADC units, corresponding to the 59.54~keV full-energy peak; $Q_{SPE}$ is the mean single-photoelectron charge; and $E_{^{241}\text{Am}}=59.54$~keV.

To determine $Q_{^{241}\text{Am}}$, the integrated-charge spectrum for each PMT was modeled as the sum of an exponential background, representing the Compton continuum, and a Gaussian full-energy peak. The Gaussian mean ($\mu$) was taken as $Q_{^{241}\text{Am}}$. As listed in Table~\ref{tab:summary}, the energy resolution, defined as $\sigma/\mu$, was $17.4 \pm 0.4\%$ for PMT1, $16.3 \pm 0.3\%$ for PMT2, and $11.3 \pm 0.1\%$ for the summed charge spectrum.

\begin{figure*}[!t]
  \centering
  \includegraphics[width=\textwidth]{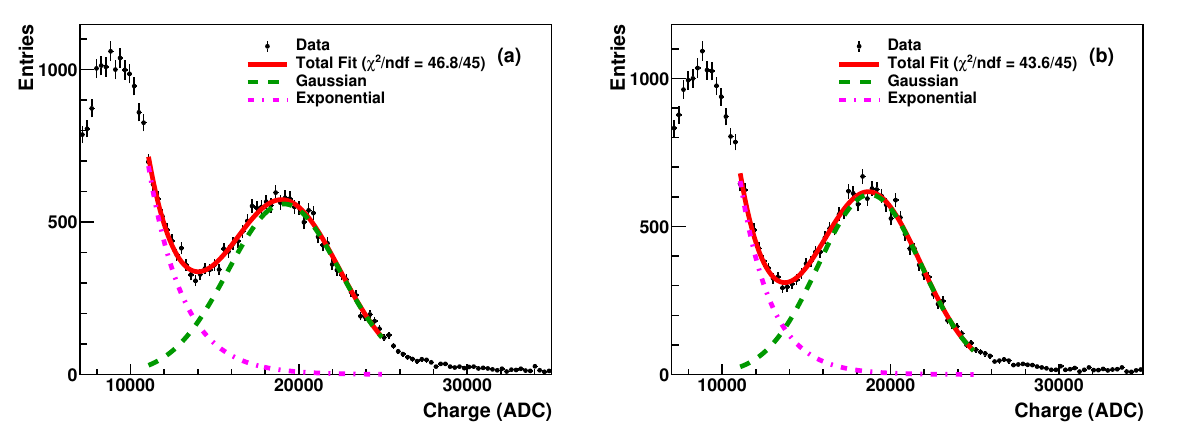}
    \caption{Integrated-charge spectra recorded with an $^{241}$Am source for (a) PMT1 and (b) PMT2. Black circles show the data. The solid red line is the total fit, consisting of a Gaussian full-energy peak (green long-dashed) and an exponential background (magenta dash-dotted).}
  \label{fig:energy}
\end{figure*}

\begin{figure*}[!tbh]
  \centering
  \includegraphics[width=\textwidth]{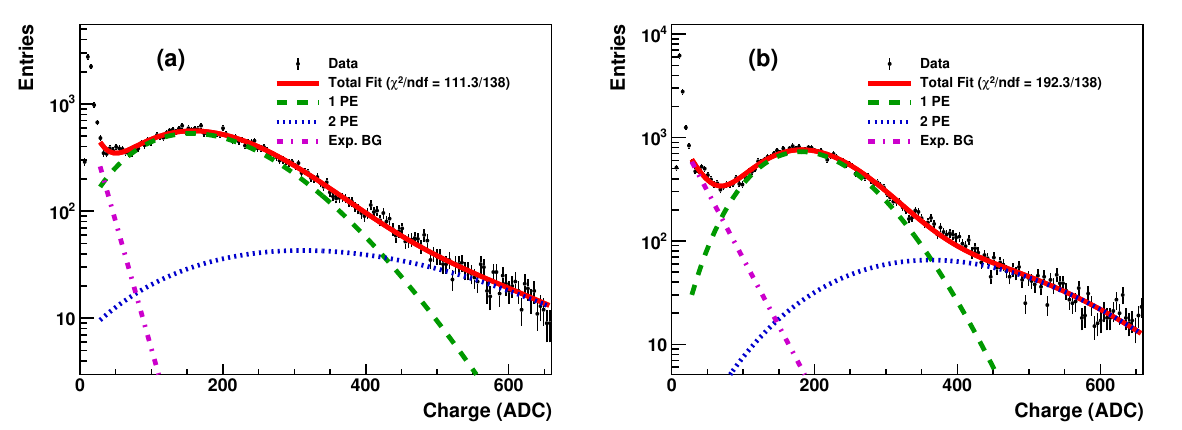}
  \caption{Single-photoelectron charge distributions extracted from the late-time tail ($t>2.53~\mu$s) for (a) PMT1 and (b) PMT2. Black circles show the data. The total fit (solid red) consists of 1-photoelectron (green long-dashed) and 2-photoelectron (blue dotted) continuous Poisson components and an exponential background (magenta dash-dotted).}
  \label{fig:spe}
\end{figure*}

A cluster-based waveform analysis was used to estimate the single-photoelectron charge ($Q_{SPE}$)~\cite{lee2006first}. A cluster was defined as a contiguous set of FADC samples above the baseline threshold, corresponding to the localized arrival of one or more photoelectrons.

To isolate predominantly single-photoelectron signals, clusters were selected only from the late-time region ($t>2.53~\mu$s), corresponding to more than 130~ns after the trigger. Events were also restricted to an approximately $3\sigma$ interval around the $^{241}$Am full-energy peak ($15{,}000 < \text{Charge} < 25{,}000$~ADC). Approximately 2.3\% of the total scintillation light, or about 4.5 photoelectrons, remained within the 5.63~$\mu$s tail window. This corresponds to an average inter-cluster interval of approximately 1.2~$\mu$s and an estimated cluster pileup probability of approximately 2\% under a Poisson assumption.

Broad electronic-noise clusters were rejected using a shape-based selection that required the ratio of cluster height to cluster width to exceed 1.5. The resulting charge distribution was fitted with a model containing continuous Poisson components for one and two photoelectrons together with an exponential background. The mean of the one-photoelectron component was taken as $Q_{SPE}$.

\begin{table}[htbp]
\centering
\caption{Energy Resolution and Effective Light Yield of the t-Stilbene Detector}
\label{tab:summary}
\begin{tabular}{@{}lccc@{}}
\toprule
\textbf{Parameter} & \textbf{PMT1} & \textbf{PMT2} & \textbf{PMT1+PMT2} \\ \midrule
Resolution ($\sigma/\mu$) (\%)       & $17.4 \pm 0.4$     & $16.3 \pm 0.3$     & $11.3 \pm 0.1$     \\
LY (NPE/keV)          & $1.76 \pm 0.02$    & $1.61 \pm 0.01$    & $\mathbf{3.37 \pm 0.02}$ \\ \bottomrule
\end{tabular}
\end{table}

As summarized in Table~\ref{tab:summary}, the effective light yields were $1.76 \pm 0.02$~NPE/keV for PMT1 and $1.61 \pm 0.01$~NPE/keV for PMT2. The summed response therefore yielded $3.37 \pm 0.02$~NPE/keV, equivalent to 3370~NPE/MeV. The Hamamatsu R12669 PMTs have a nominal peak quantum efficiency (QE) of approximately 35\% near 420~nm. At the measured emission peak of 380~nm, the QE is estimated to be approximately 30\%. Dividing the measured photoelectron yield by this QE gives a lower-bound estimate of approximately 11,000 photons/MeV under ideal light-collection conditions; optical collection losses would increase the inferred intrinsic yield. This value is comparable to the typical light yields of conventional liquid and organic scintillators, approximately 10,000 photons/MeV \cite{galunov2013}, and supports further evaluation of this t-stilbene crystal for low-mass dark matter searches.

\section{Conclusion}
\label{sec:conclusion}
We characterized the optical and scintillation properties of a solution-grown t-stilbene crystal. The photoluminescence spectrum exhibits a dominant emission peak near 380~nm. The scintillation waveform is described by a triple-exponential model with fraction-weighted decay constants of 8.3~ns, 20.1~ns, and 74.2~ns for the fast, medium, and slow components, respectively. The fast component is longer than typical intrinsic values reported in the literature, likely because the measured waveform includes the response of the electronics and data acquisition chain.

A single-photoelectron calibration using an $^{241}$Am source yielded an effective light output of $3.37 \pm 0.02$~NPE/keV. Using the estimated PMT quantum efficiency at the emission wavelength, this photoelectron yield implies a lower-bound intrinsic light yield of approximately 11,200 photons/MeV under ideal light-collection conditions. These results demonstrate promising light-output performance and motivate further detector characterization.

Future work will evaluate the pulse-shape discrimination performance and measure the proton-recoil quenching factor using a neutron generator. After these laboratory measurements, the detector will be deployed at the underground Yemilab facility \cite{park2024} to measure environmental backgrounds and assess the feasibility of t-stilbene as a target material for low-mass dark matter searches.

\section*{Acknowledgment} 
We thank the IBS Research Solution Center (RSC) for providing high performance computing resources. 
This work is supported by: the Institute for Basic Science (IBS) under project code IBS-R016-A1,  NRF-2021R1A2C3010989, NRF-2021R1A2C1013761, RS-2024-00356960, RS-2025-25442707 and RS-2025-16064659, Republic of Korea. 
\newpage

\end{document}